\documentclass[twocolumn,showpacs,superscriptaddress,preprintnumbers,amsmath,amssymb,epsfig,floatfix,prb]{revtex4-1}
\usepackage{amssymb}
\usepackage{graphicx}
\usepackage{bm}
\usepackage{color}
\usepackage{xcolor}
\definecolor{green}{HTML}{009900}
\begin{document}

\title{Manipulation of edge magnetism in hexagonal graphene nanoflake}
\author{Mukul Kabir}
\affiliation{Indian Institute of Science Education and Research, Pune 411008, India.}
\author{T. Saha-Dasgupta}
\affiliation{S. N. Bose National Centre for Basic Sciences, Kolkata 700098, India.}
\date{\today}                                         

\begin{abstract}
We explore possible ways to manipulate the intrinsic edge magnetism in hexagonal graphene nanoflake with zigzag edges, using density functional theory supplemented with on-site Coulomb interaction. The effect of carrier doping, chemical modification at the edge, and finite temperature on the edge magnetism has been studied. The magnetic phase diagram with varied carrier doping, and on-site Coulomb interaction is found to be complex. In addition to the intrinsic antiferromagnetic solution, as predicted for charge neutral hexagonal nanoflake, fully polarized ferromagnetic, and mixed phase solutions are obtained depending on the doped carrier concentration, and on-site Coulomb interaction. The complexity arises due to the competing nature of local Coulomb interaction and carrier doping, favoring antiferromagnetic and ferromagnetic coupling, respectively. Chemical modification of the edge atoms by hydrogen leads to partial quenching of local moments, giving rise to a richer phase diagram consisting of antiferromagnetic, ferromagnetic, mixed, and nonmagnetic phases. We further report the influence of temperature on the long-range magnetic ordering at the edge using {\it ab initio} molecular dynamics. In agreement with the recent experimental observations, we find that temperature can also alter the magnetic state of neutral nanoflake, which is otherwise antiferromagnetic at zero temperature. These findings will have important implications in controlling magnetism in graphene based low dimensional structures for technological purpose, and in understanding varied experimental reports.
\end{abstract}

\maketitle {}

\section{Introduction} 

Graphene has attracted considerable attention due to its unique physical properties, and plausible applications in novel electronic devices.~\cite{Nature.6.183,Katsnelson200720} In particular, the unconventional magnetism in graphene based nanostructures is interesting, which may be utilized for spintronics applications.~\cite{RPP.2010, NatureMaterials.11,PhysRevLett.100.047209,PhysRevB.81.165409,NaturePhysics.8} Such graphene nanostructures have been fabricated experimentally,~\cite{doi:10.1021/nl048111+,PhysRevLett.99.166804,NatureChem.5,PhysRevB.88.125433,PhysRevLett.101.126102,1367-2630-12-4-043028,doi:10.1021/nn203381k,Geng16042012} and interestingly magnetism  has been reported in nanographene,~\cite{PhysRevLett.84.1744,Enoki20091144} disordered graphite,~\cite{PhysRevB.66.024429,PhysRevLett.91.227201} and grain boundaries in highly oriented pyrolytic graphite.~\cite{Nature.Physics.5}

The electronic structure of graphene nanostructure is different from the infinite graphene due to the edge effect.~\cite{PhysRevB.54.17954} The existence of edge and its nature influences the properties of carbon $\pi$ electrons in a non trivial manner. These electrons in graphene can be represented as mass-less Dirac fermion in a bipartite lattice. The two graphene sublattices give rise to two pseudo-spins with two degrees of freedom. For a zigzag edge, the atoms at a particular edge belong to either of the two sublattices, which breaks the pseudo-spin symmetry of the Dirac fermion.  In contrast, the situation in armchair edge is very different, which contain atoms from both sublattice, and the pseudo-spins are always paired. In an alternative view, the zigzag edges give rise to localized nonbonding $\pi$ states at the Fermi level. This is in contrast with the armchair edges, which has no such edge states. This is analogous with the distinction between Kekule$^{\prime}$ and non-Kekule$^{\prime}$ polycyclic hydrocarbon molecules.~\cite{Enoki20091144}  These theoretical predictions have been confirmed experimentally via scanning tunneling microscopy and scanning tunneling spectroscopy.~\cite{PhysRevB.71.193406,PhysRevB.73.125415,PhysRevB.73.085421,Nature.Physics.7}

The appearance of half-filled flat band at the Fermi level gives rise to localized moment at the zigzag edge,~\cite{PhysRevB.54.17954} and long-range magnetic order have been predicted theoretically.~\cite{doi:10.1143/JPSJ.65.1920, PhysRevLett.100.047209,PhysRevLett.102.227205,PhysRevLett.99.177204,jcp.128} The nature of  long-range coupling between the edges is dictated by whether the sites at the neighboring edges belong to the same or different sublattice of graphene. For example, atoms in a particular edge of regular hexagonal nanographene belong to the same sublattice, while atoms in the alternate edges belong to different sublattices. Thus, the intra-edge coupling is ferromagnetic, and inter-edge coupling is antiferromagnetic, which gives rise to a fully compensated ferrimagnetic solution with zero magnetic moment.~\cite{PhysRevLett.99.177204} This is in accordance with the Lieb's theorem.\cite{PhysRevLett.62.1201}  Thus, it is worth exploring the possible ways to manipulate this intrinsic antiferromagnetic order toward an uncompensated magnetic order with net magnetic moment, which could be utilized in spintronics applications. Furthermore, recent experimental study predicts that magnetism in zigzag nanographene is controversial with varied reports of ferromagnetism,~\cite{PhysRevLett.98.187204,PhysRevB.81.245428} antiferromagnetism, and diamagnetism, and their coexistence.~\cite{Scientific.Reports.3} Therefore, the experimental situtaion concerning magnetism in graphene nanostructures remains debated, and is far from being fully understood. In this context, it will be important to study the plausible influence of various external perturbations on the intrensic magnetism, which may arise depending on the experimental conditions.

With the above motivations, here we investigate the influence of carrier doping, chemical modification of the edge, and temperature on the long-range magnetic order in hexagonal zigzag nanographene. Present first-principles density functional theory based calculations show that the compensated ferrimagnetic solution with zero total moment evolves into a different magnetic phase with finite total moment upon carrier doping. This opens up a possible emergence of polarized solution with parallel spin alignment at the edges, which will be of technological importance. We also find that chemical modification of the edges by mono-hydrogenation has important implication on the long-range magnetic coupling.  Further, we show that such change in magnetic coupling can also be induced by temperature, which is in agreement with the recent experimental observations.~\cite{Scientific.Reports.3}  We believe the present study will contribute to the understanding of edge magnetism in nanographene. 

\section{Computational Details} 
Calculations are carried out using density functional theory (DFT) implemented in the Vienna {\it Ab-initio}  Simulation Package,~\cite{PhysRevB.47.558,PhysRevB.54.11169} with the projector augmented wave pseudopotential.~\cite{PhysRevB.50.17953} For exchange-correlation functional we use the Perdew-Burke-Ernzerhof (PBE) form of generalized gradient approximation (GGA),~\cite{PhysRevLett.77.3865} and a plane-wave cutoff of 800 eV is used.  To consider the strong electron correlation, the on-site Coulomb interaction is added to the PBE functional (DFT+$U$) within Dudarev's approach.\cite{PhysRevB.57.1505} Reciprocal space integrations are carried out at the $\Gamma$ point. Simple cubic supercells with periodic boundary condition are used, where two neighboring flakes are kept separated by at least 12 {\AA} vacuum space. This ensures the interaction between the images to be negligible. All the structures are optimized until the forces on each atom become less than 5 $\times$ 10$^{-3}$ eV/\AA.

We perform {\it ab initio} molecular dynamics (AIMD) simulation, where the canonical ensemble is employed with target temperature of 100, 200 and 300K, maintained by a Nos\'e-Hoover thermostat.~\cite{PhysRevB.33.339, PhysRevA.31.1695} We consider 1 fs time step, and the sample is equilibrated for 2 ps at the target temperature. The trend in free energy is then observed for next 10 ps.  At each time step the energy is converged to 10$^{-5}$ eV prior to force evaluation. 

\section{Results}
To establish our methodology, we first study the stability of edge magnetism in bare zigzag nanoflake, which is theoretically known to provide fully compensated ferrimagnetic solution.\cite{PhysRevLett.99.177204} Next, we investigate the plausible ways to manipulate the edge magnetism in hexagonal nanographene, which is the main goal of the present study. In this regard, we study the effects of (a) carrier doping, (b) chemical modification of the edges, and (c) application of finite temperature on the edge magnetism. 

\subsection{Edge magnetism in hexagonal nanoflake}

\begin{figure}[t]
\begin{center}
\rotatebox{0}{\includegraphics[width=0.5\textwidth]{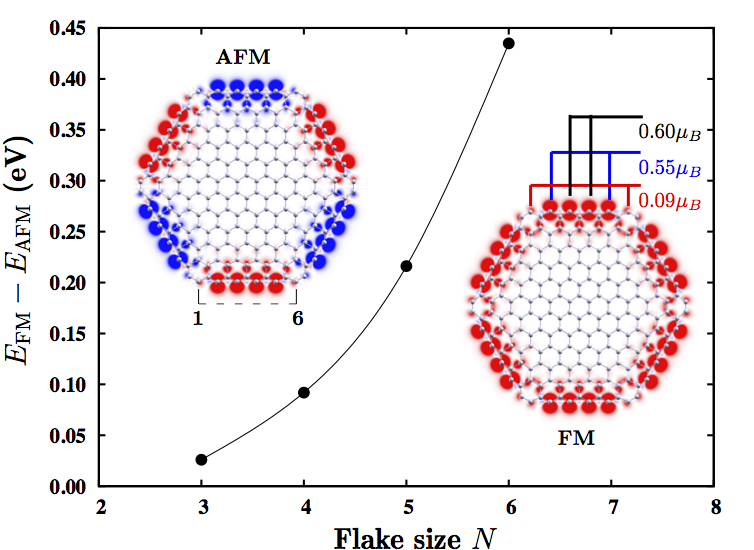}}
\end{center}
\caption{(Color online) The energy difference between the AFM ground state and the excited FM state for charge neutral bare flakes increases with increasing flake size. Energies are calculated using PBE functional without considering the local Coulomb interaction.  The AFM and FM magnetization densities are shown for $N$ = 6 flake.  Up and down densities are indicated with red (light gray) and black (dark gray) colors, respectively.  The carbon atoms at the edge has substantial localized magnetic moments, which show a distribution in their values among the inequivalent sites as shown for the FM case.}
\label{figure:1}
\end{figure}

We start our discussion by revisiting the electronic and magnetic structure of charge neutral bare hexagonal graphene nanoflakes of various sizes. The size of the nanoflake is denoted by the number of six-member rings at each edge $N$ (Fig.~\ref{figure:1}). Hexagonal nanoflake with zigzag edges has equal number of atoms belonging to two different sublattices resulting into $S$=0 ground state according to Lieb's theorem.~\cite{PhysRevLett.62.1201} The GGA calculations predict substantial magnetic moments at the edge sites for all the flakes studied here. These moments at a particular edge are found to be parallel, which are aligned antiparallely with the neighboring edge sites (Fig.~\ref{figure:1}). This gives rise to an antiferromagnetic (AFM) or compensated ferrimagnetic state, and is in agreement with the established knowledge.~\cite{PhysRevLett.99.177204}  Moreover, local moments at the edge show a distribution in their values, as two adjacent zigzag edges are connected by an armchair defect. The calculated local moment distribution, and the magnetization densities are shown in the inset of Fig.~\ref{figure:1} for $N$=6 nanoflake. The nature of the distribution, with the largest moment residing at the central site of the edge, is also in agreement with the previous observations,\cite{PhysRevLett.99.177204} arising due to the distribution in electron localization along the zigzag edge. Similar distribution of electron localization has been also predicted using Dirac equation in zigzag edges.~\cite{PhysRevB.84.245403}  The above results are in agreement with the previous calculations,~\cite{PhysRevLett.99.177204,PhysRevB.84.245403} and therefore validate the methodology adopted in the present study.  Additionally, we find that there exists a solution where all the edge sites are parallely aligned, namely the ferromagnetic (FM) state, which lies higher in energy than the AFM ground state for all sizes. The energy difference between the FM excited state ($E_{\rm FM}$) and the AFM ground state ($E_{\rm AFM}$) increases with increasing flake size (Fig.~\ref{figure:1}). Further, we calculate the energetics of the nonmagnetic solutions ($S$=0), which are found to be much higher (3--8 eV depending on the size of the flake) compared to the corresponding AFM ground state. This suggests the magnetic solutions to be robust. Since $E_{\rm FM} - E_{\rm AFM}$ is found to  increase with increasing flake size, it is expected that achieving polarized solution will be comparatively easier for smaller flakes. Thus, in our subsequent discussion, we focus on the specific case of $N$=3 nanoflake.

\begin{figure}[t]
\begin{center}
\rotatebox{0}{\includegraphics[width=0.5\textwidth]{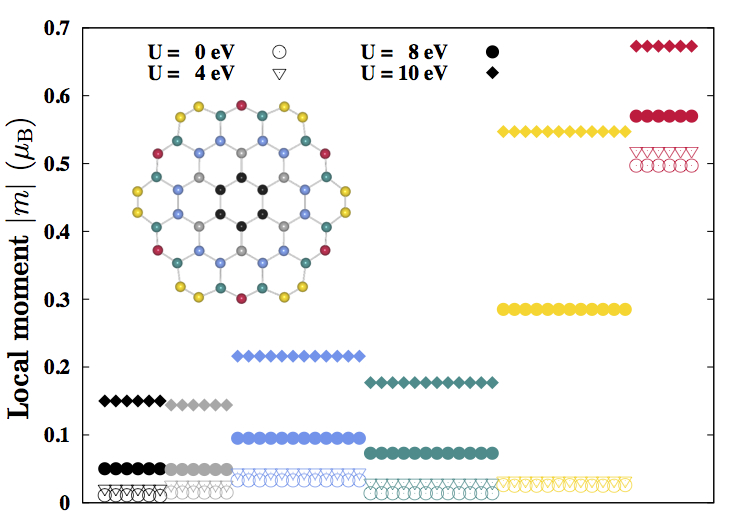}}
\end{center}
\caption{(Color online) Distribution of local magnetic moments at various inequivalent carbon sites for $N$=3 nanoflake in the AFM ground state. The results are shown for selected on-site Coulomb interaction $U$. The magnetic moments at inequivalent sites are marked in different colors, and the same coloring scheme is followed as in the inset. For a particular edge, two inequivalent edge atoms have different local moments, and the difference decreases with increasing on-site Coulomb interaction.}
\label{figure:2}
\end{figure}

The strength of electron-electron correlation in carbon based molecules, and graphene is a matter of discussion in recent literature.~\cite{RevModPhys.81.109, jcp.1561,PhysRevLett.54.1844, PhysRevLett.56.1509, PhysRevLett.106.236805, 0295-5075-19-8-007, PhysRevB.72.085123, Nature.464} The bare on-site Coulomb interaction in benzene was estimated to be very high (17 eV),~\cite{jcp.1561} while this was estimated to be 10 eV for polyacetylene molecule.~\cite{PhysRevLett.54.1844, PhysRevLett.56.1509} Recent study argued the electron-electron correlation in graphene to be substantially large.\cite{PhysRevLett.106.236805} The local Coulomb interaction was predicted to be smaller than a critical value for which a fully compensated antiferromagnetic solution is favorable,~\cite{0295-5075-19-8-007, PhysRevB.72.085123} but close to 9.3 eV which favors spin-liquid solution.~\cite{Nature.464} Following this proposal, we repeat calculations by adding the on-site Coulomb interaction to the PBE-GGA exchange-correlation functional (DFT+$U$), and vary  $U$ from 0 (GGA) to 10 eV. Fig.~\ref{figure:2} shows the variation of local magnetic moment at various inequivalent carbon sites for $N$=3 flake for various $U$ values.  The on-site Coulomb interaction enhances the localization of C-$p$ states, and thus the local magnetic moment at individual edge sites increases with increasing $U$. For $U\geqslant$ 4 eV, the atoms at the `core' also attain a finite local moment, which increases with increasing $U$. Nevertheless, the magnetic moment of the edge atoms are found to be substantially higher than that of the core atoms, indicating higher localization of the edge states. For example, the maximum local magnetic moment of core atoms is found to be $\sim$0.20$\mu_{B}$ for $U$=10 eV, which is much smaller than the maximum moment at the edge, 0.68$\mu_{B}$. This ensures the survival of edge magnetism even for large on-site Coulomb interaction. It is important to point out here that all calculations involve structural optimization of the flake, and the on-site Coulomb interaction is found to have significant influence on the optimized structure. The average C--C bond length is found to increase with increasing $U$ (see Supplementary Information).~\cite{supple} In all cases, the average C--C bond length at the edge is found to be shorter ($\sim$0.08 \AA)~ than those in the core.~\cite{supple}

\subsection{Manipulation of long-range magnetic order through carrier doping}

In this subsection we discuss the effect of carrier (hole or electron) doping on the edge magnetism of nanoflake. The neutral flake has one $\pi$ electron at each carbon site,  and within half-filled single-band Hubbard model, the existence of ferromagnetic phase in infinite bipartite lattice has been proposed at the Nagaoka limit -- single electron/hole doping, and infinite on-site Coulomb interaction.~\cite{PhysRev.147.392} This motivated us to investigate the effect of carrier doping on the intrinsic AFM state of the undoped flake. However, the present situation is rather different from the Nagaoka limit in many ways -- (i) finite size of the nanoflake, (ii) finite value of on-site Coulomb interaction, (iii) multi-band treatment within DFT, and (iv) structural modification of the finite sized flake.~\cite{supple} 

\begin{figure}
\begin{center}
\rotatebox{0}{\includegraphics[width=0.45\textwidth]{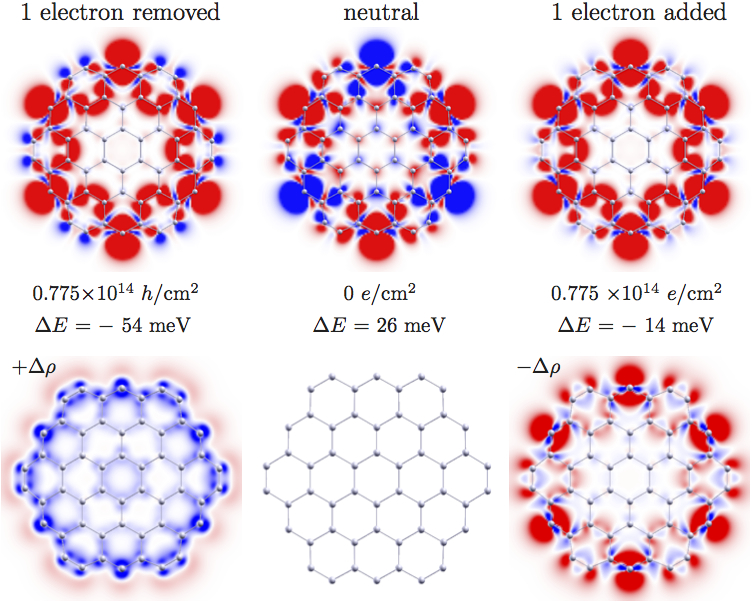}}
\end{center}
\caption{(Color online) The GGA magnetization densities for bare nanoflakes with change in carrier concentration; charge neutral (upper middle), and single hole (upper left) and single electron doped (upper right). Red  (light gray) and blue (dark gray) color represent up and down magnetization densities, respectively. Calculated energy difference, $\Delta E$ = $E_{\rm FM} - E_{\rm AFM}$ shows a change in magnetic coupling due to single electron/hole doping. The bottom panel shows the charge density difference between the carrier doped and charge neutral nanoflake. Note that the doped carrier (electron or hole) is distributed only at the edges.}
\label{figure:3}
\end{figure}

\begin{figure}[t]
\begin{center}
\rotatebox{0}{\includegraphics[width=0.47\textwidth]{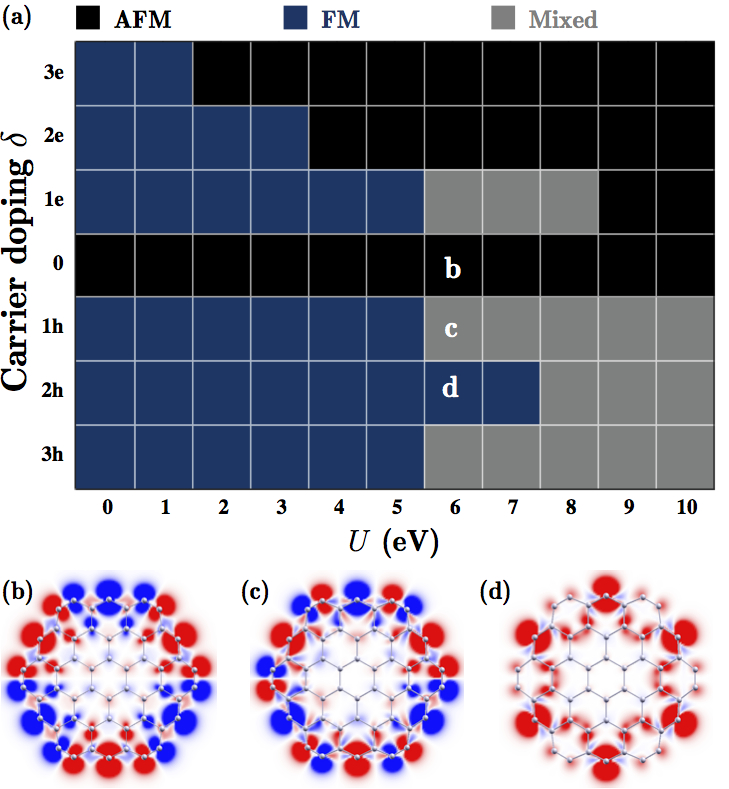}}
\end{center}
\caption{(Color online) (a) Magnetic phase diagram for bare $N$=3 nanoflake with varied on-site Coulomb repulsion and carrier density, as obtained within DFT+$U$ calculations. The antiferromagnetic, ferromagnetic and the mixed state solutions are indicated with different color (gray) shades.  We observe an emergence of FM ordering due to carrier doping. Representative spin densities for (b) AFM, (c) mixed, and (d) FM configurations are shown for  $U$=6 eV.}
\label{figure:4}
\end{figure}

Within the conventional PBE-GGA calculation without considering on-site Coulomb interaction, we indeed find that the long-range magnetic coupling between the hexagonal edges becomes ferromagnetic due to single hole/electron doping (Fig.~\ref{figure:3}). This corresponds to a carrier doping of 0.775 $\times$ 10$^{14}$ $h (e)$ cm$^{-2}$. Such high carrier density has been experimentally achieved (0.2 $\times$ 10$^{14}$ cm$^{-2}$).\cite{Nat.Comm.4}  The calculated energy difference $\Delta E$ shows that the relative stabilization of the FM state is greater for hole doping compared to that of the electron doping, which points toward an electron-hole asymmetry. The deviation from the particle-hole symmetry, which is the feature of infinite bipartite lattice, is due to the finite size of the nanoflake, and the corresponding deviation from perfect hexagonal lattice upon structural optimization. The FM solution remains the ground state upon further increase in doping (Fig.~\ref{figure:4}). The energy differences $\Delta E$ between the FM ground state and the first excited AFM states vary between $-$50 to $-$150 meV depending on the type and concentration of doping. Similar to single carrier doping, at high doping concentrations, the FM solutions are comparatively more stable for hole doped cases than for electron doped cases. In order to keep the doped carrier density close to the experimentally achieved limit, we did not consider more than three hole/electron doping. We find the emergence of inter-edge FM coupling for both electron and hole doping to be driven by the intrinsic electronic structure. We observe that the density of state at the Fermi level is higher in the case of AFM solution compared to that of the corresponding FM solution, which indicate higher stability of the FM solution. Charge density difference between the charge neutral and the doped cases (Fig.~\ref{figure:3}) shows that the added electron or hole is distributed at the six edges, justifying the edge states being responsible for the switching of magnetic coupling.  Different density distribution of electron and hole (Fig.~\ref{figure:3}) further emphasizes the particle-hole asymmetry. The calculated total magnetic moment for $N$=3 nanoflake in the ferromagnetic state is found to be 6, 7, and 5$\mu_{B}$ for neutral, hole, and electron doped cases, respectively. As discussed earlier and shown in Fig.~\ref{figure:2}, for a particular edge the localized moments have a distribution, and the largest moment is found to be 0.52, and 0.40$\mu_B$ for single hole, and single electron doped cases, respectively.  These moments are comparable to the AFM solution for the neutral flake (0.5$\mu_B$), which suggests that carrier doping primarily modulates the sign of the magnetic coupling, rather than influencing the individual magnetic moments. The change in long range magnetic order due to carrier doping is also found to be accompanied by opening (closing) of gap between the highest occupied and lowest unoccupied molecular level for majority (minority) channel. For example, the AFM ground state has 1.40 (1.40) eV gap for the majority (minority) channel, which becomes 2.01 (0.13) and 2.00 (1.19) eV due to one electron and hole doping, respectively. This suggests that carrier doping, in addition to influencing the magnetic ordering, should also affect the spin polarized transport.     

We extend our study to consider strong Coulomb interaction proposed for graphene nanostructures,\cite{RevModPhys.81.109, PhysRevLett.106.236805} and also vary the doped carrier density.  The calculated magnetic phase diagram is shown in Fig.~\ref{figure:4} with varying carrier doping ($\delta=$ 3$h$--3$e$), and on-site Coulomb $U$ (0--10 eV). For finite $U$, the charge neutral flake remains antiferromagnetic, and the stability of AFM solution increases with increasing $U$ as the correlation increases. Moreover, the FM solution becomes unstable at large $U$ ($\geqslant$ 9 eV). These AFM solutions have usual ferromagnetic intra-edge coupling, and antiferromagnetic inter-edge coupling, which is shown for $U$=6 eV [Fig.~\ref{figure:4}(b)].  The situation for finite $U$ with carrier doping is found to be very complex, and driven by the interplay between various different factors. We observe a strong asymmetry in the phase diagram between electron and hole doping. We find that for $U>$1 eV, the asymmetry in the phase diagram becomes qualitative in terms of predicting different magnetic ground states for hole and electron doping. Such asymmetry for $U$=0 eV was only quantitative -- both hole and electron doping predicts FM structure but with different $\Delta E$. It is already mentioned that the on-site Coulomb interaction influences the optimized structure -- average bond length increases with increasing $U$.~\cite{supple} In addition, the nature (electron or hole) and concentration of carrier doping are found to have significant effect on the optimized structure through angular distortion at the corners of the edges, which are armchair defects.~\cite{supple} While the charge neutral flake exhibits no significant angular distortion for all $U$ studied here, this angular distortion increases with increasing electron doping, and further increases with $U$. In contrast, such distortion for hole doping is found to be negligibly small.  A detailed structural analysis is presented in the Supplementary Information.~\cite{supple} This correlation and doping dependent structural distortion makes the behavior of electron doped flakes markedly different compared to that of the hole doped ones. Above a threshold value of this angular distortion, the edge geometry deviates significantly from the zigzag structure. Such angular distortion reduces the local moments at all the sites on the distorted edge. These moments become comparable in magnitude, which makes the nearest-neighbour AFM correlation even stronger. This gives rise to an antiferromagnetic solution for sufficiently high electron doping and large $U$ In the following, we discuss the hole and electron-doped part of the phase diagram in more detail.

First we discuss the scenario for hole doping. For intermediate $U$ ($\leqslant$ 5 eV), the inter-edge coupling becomes ferromagnetic for hole doping at all concentrations studied here. The first excited states are always found to be AFM, and $\Delta E$ depends on the strength of carrier doping and on-site Coulomb $U$, which ranges between $-$50 and $-250$ meV. In contrast, for large $U$ ($\geqslant$ 6 eV) a new type of magnetic state (mixed) emerges [Fig.~\ref{figure:4}(c)]. In this state, both inter-edge and intra-edge magnetic couplings do not follow a definite order, and are very different from either of the AFM [Fig.~\ref{figure:4}(b)] and FM [Fig.~\ref{figure:4}(d)] solutions. Such mixed magnetic structure may be considered similar to a `paramagnetic' state. The emergence of FM and mixed states are due to the competing effects of $U$ and hole doping -- while $U$ favors the AFM coupling, hole doping tries to make the inter-edge coupling ferromagnetic.  Below a critical value of $U$ (6 eV), the effect of hole doping dominates over the Coulomb interaction, while for large $U$, the competition makes the mixed magnetic structure favorable. For mixed ground states, the corresponding first excited states are found to be AFM, with $\sim$ 100--300 meV energy difference. We remind here that the angular distortion remains negligible for the entire range of Coulomb interaction and hole concentration.

In contrast to hole doping, the situation for electron doping is more complex as it introduces angular distortion at the corner of the edges. As mentioned, this distortion critically depends on $U$ and doping level.~\cite{supple} We observe a clear correspondence between the structural distortion and the obtained magnetic structure. If the angular distortion is negligible, the electron doping promotes FM inter-edge coupling.  On the other hand, if the correlation and doping driven angular distortion is significant, the AFM structure is found to be stabilized. In such cases, both the Coulomb interaction $U$ and the structural distortion favors AFM coupling.  We also observe the appearance of mixed states for single electron doping in the range $U$=5--8 eV. We did not find significant structural distortion for these structures, and thus the emergence of mixed phase is governed by the competition between $U$ and electron doping. For large $U$, the ground state is found to be either AFM or mixed state, and the corresponding first excited state becomes mixed or AFM, respectively.  Stabilization of FM states in such cases is not achieved due to the dominance of correlation and structural distortion driven antiferromagnetism. Although, the present situation and theoretical treatment within DFT is very different from the Nagaoka limit of single-band Hubbard model, the isolated regions of FM and mixed magnetic phases have been reported via an emergence of Nagaoka state in infinite hexagonal lattice at large on-site Coulomb interaction.~\cite{ANDP:ANDP19955070405}

\subsection{Effect of hydrogenation on long-range magnetic order}

\begin{figure}[t]
\begin{center}
\rotatebox{0}{\includegraphics[width=0.5\textwidth]{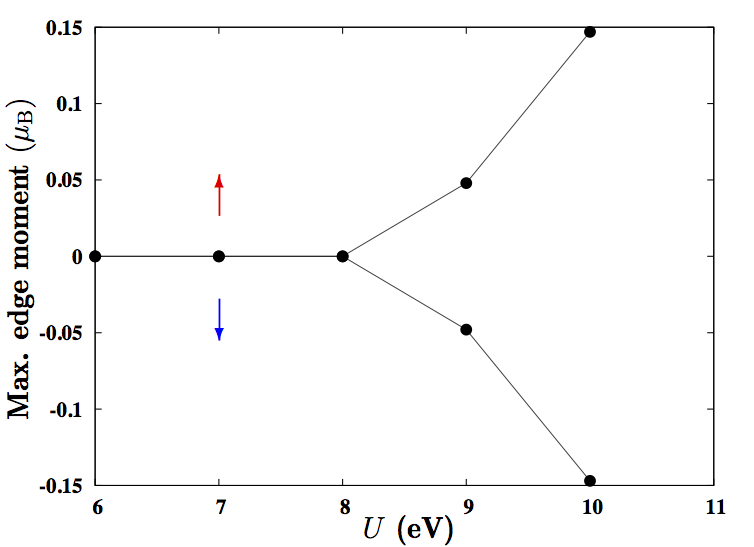}}
\end{center}
\caption{(Color online) Maximum localized moment at the edge with varied on-site Coulomb interaction for charge neutral $N$=3 H-nanoflake. Local moments for sites belonging to both sublattice are shown, which are equal in magnitude and AFM coupled. Due to H-passivation at the edges, the local moments are quenched, and for $U\leqslant$8 eV the charge neutral flake becomes nonmagnetic.}
\label{figure:5}
\end{figure}

We next discuss the effect of chemical modification on the magnetic phase diagram. Edge states are susceptible to chemical modification, and hydrogenated nanoflakes have been earlier studied theoretically,~\cite{doi:10.1021/nl072548a} and experimentally.~\cite{PhysRevB.71.193406} Recently, H- and F-passivated charge neutral graphene flakes with different geometry has been studied. The study showed that the calculated gap between the highest occupied and lowest unoccupied molecular orbital strongly depends on the number of carbon atoms in the flake, with a possibility of half-metallic solution at certain sizes. This suggests that chemical modification of edge states can have profound effect on its electronic structure.~\cite{10.1063/1.4865414} For the present study, the dangling bonds at the edge are terminated with single hydrogen on the plane of the flake. For charge neutral nanoflake we find, within PBE-GGA calculations, that the local moments at the edge sites are completely quenched due to hydrogenation, and the nanoflake becomes nonmagnetic. This is expected since the addition of H creates C--H bond at the edges, which passivates the dangling bond responsible for magnetism. However, such GGA-PBE calculations does not account for the strong electron-electron correlation, which is argued to be important for graphene and derived nanostructures.~\cite{RevModPhys.81.109,PhysRevLett.106.236805} Thus, we carried out calculations within the DFT+$U$ framework for charge neutral H-nanoflakes, and varied the on-site Coulomb interaction between 0--10 eV. Beyond a critical Coulomb interaction ($U_{\rm cr} \geq$ 8 eV), atoms at the edge for neutral H-nanoflake regain finite local moment (Fig.~\ref{figure:5}), and the nanoflake becomes an usual compensated ferrimagnet, with ferromagnetic intra-edge coupling and antiferromagnetic inter-edge coupling, as found in the case of charge neutral bare nanoflakes.  

\begin{figure}[t]
\begin{center}
\rotatebox{0}{\includegraphics[width=0.47\textwidth]{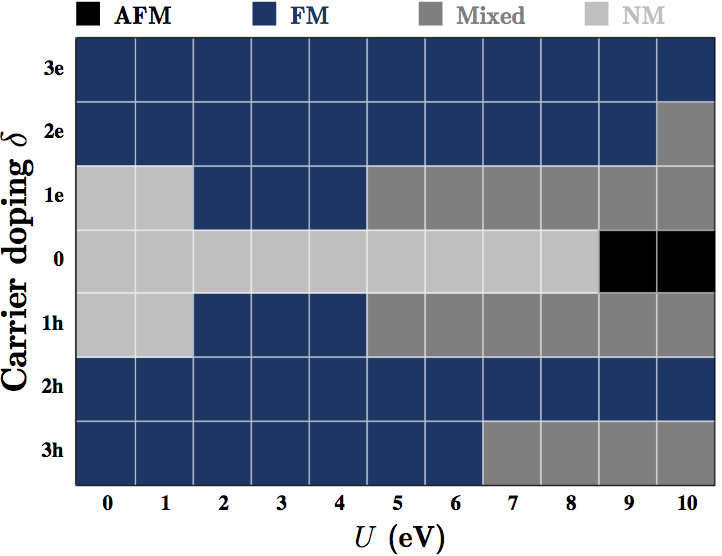}}
\end{center}
\caption{(Color online) Magnetic phase diagram with varied on-site Coulomb interaction and charge doping for $N$=3 H-nanoflake. Different magnetic phases includes antiferromagnetic, ferromagnetic, nonmagnetic, and mixed state solution, which are indicated with different color (gray) shades.}
\label{figure:6}
\end{figure}

As in the case of bare nanoflake, we studied the magnetic phase diagram with varied carrier concentration, and on-site Coulomb interaction. The magnetic $\delta-U$ phase diagram for H-nanoflake is shown in Fig.~\ref{figure:6}, which is richer, and significantly different from the same obtained for bare nanoflake (Fig.~\ref{figure:4}). In addition to the three different magnetic phases (AFM, FM and mixed) for bare nanoflake, the magnetic phase diagram for H-nanoflake contains a non-magnetic (NM) phase, where individual sites have zero local moment. Moreover, compared to bare nanoflake, the asymmetry in the phase diagram in the context of predicted magnetic phase for hole and electron doping is less prominent. Below $U\leqslant$6 eV, the phase diagram is completely symmetric, and predicts the same magnetic solution upon hole and electron doping at the same density. However, for $U\geqslant$7 eV, we find that the magnetic ground state in some cases are different for hole doping than that of the electron doping at same density. Interestingly, compared to the bare flake [Fig.~\ref{figure:4}], we find that for a wide range of $U$ and doping level the FM solution becomes favorable.  The AFM solution emerges as ground state only for charge neutral H-flake at $U\geqslant$9 eV.  This clearly establishes that stabilization of FM phase is easier in H-nanoflake compared to the corresponding bare one. For carrier doped H-flake, we find stable NM, mixed, and FM solutions, while the AFM solutions are not stabilized. The calculated energy difference between the magnetic ground state and the corresponding NM solution varies within 10--100 meV range.

Similar to the bare nanoflake, the average bond length in the optimized structure is found to increase with $U$, which is expected due to the increase in electron correlation. However, this is important to note that the difference in the bond length at the edge and the core is much smaller (0.01--0.02 \AA) than the bare counterparts ($\sim$0.08 \AA).~\cite{supple} Moreover, in contrast to the bare flake, for all the structures in the studied range of $U$ and doped carrier concentration the angular distortion is found to be nearly zero due to the H-passivation.~\cite{supple} The passivation-driven reduction of the local moment at the edge sites, together with reduction in structural distortion, makes the influence of $U$ less dominant in H-flake, compared to the bare flake. This causes relative strengthening of doping effect over the correlation effect, and thereby promote FM phase over a large range of $U$ and doped carrier concentration.

As discussed earlier, for charged neutral flake, H-passivation is found to completely quench the local moments, which becomes 
finite only for $U\geqslant$ 9 eV. However, surprisingly a magnetic solution (FM or mixed) appears to be favorable over the nonmagnetic solution due to carrier doping. For moderate $U$ (0--4 eV), the description of weakly correlated picture holds good, where FM solution emerges due to carrier doping, which is otherwise nonmagnetic for the undoped case. For carrier doped H-nanoflake, the GGA calculation predicts the FM ground state to be close in energy with the corresponding nonmagnetic solution. The energy difference is found to be 7 and 24 meV for two hole and two electron doped cases, respectively, which is much smaller than the bare counterparts.  An investigation of the corresponding density of states reveals that the stabilization of FM state takes place primarily by the formation of Stoner instability in the electronic structure of the carrier doped H-nanoflakes. Above $U\geqslant$ 5 eV, we observe an emergence of mixed magnetic phase similar to the mixed phase in bare nanoflake due to a close competition between the on-site Coulomb $U$, and carrier doping.   

The net magnetic moment of the H-nanoflake ($\sim$ 1$\mu_B$) is much smaller than that of the corresponding bare flake due to the weakening of the local moment at the edge through H-passivation. Nevertheless, 1$\mu_B$ moment is reasonable for device application. Considering the fact that magnetic phase diagram contains a large region of FM state, the H-flakes are worthy of investigation for probable spintronics applications. It is important to mention that the picture for larger H-flakes should be qualitatively similar. For example, we find a FM ground state with $\sim$ 3.5$\mu_B$ moment for $N$=6 H-flake upon 1.4 $\times$ 10$^{14}$ $e$ cm$^{-2}$ doping at $U$=9 eV.

\begin{figure}[t]
\begin{center}
\rotatebox{0}{\includegraphics[width=0.47\textwidth]{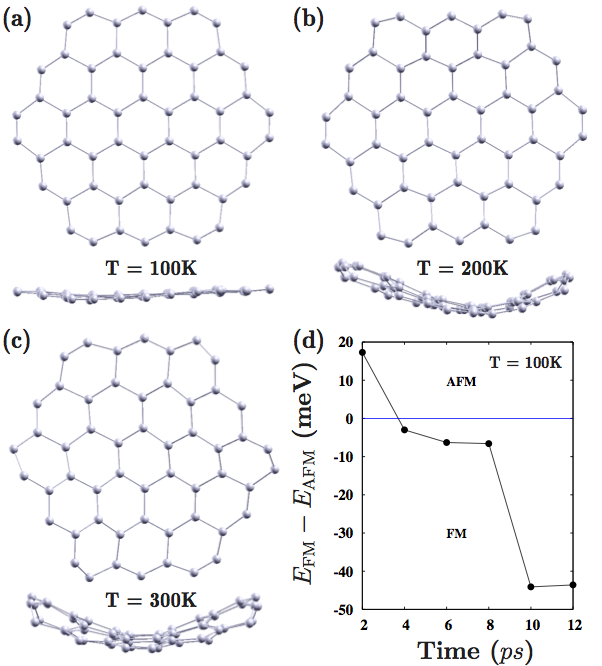}}
\end{center}
\caption{(Color online) Effect of temperature on the geometric and magnetic structure for charge neutral nanoflake. Geometric structures are shown for (a) 100K,  (b) 200K, and (c) 300K. Two dimensional structure is retained till 100K, which is severely distorted at elevated temperatures beyond 100K.   
(d) The free energy difference between the fully compensated ferrimagnetic structure and the polarized ferromagnetic structure indicate the FM structure to be the ground state at 100K, which is otherwise AFM at 0K. This result is in good qualitative agreement with the recent experimental observation.~\cite{Scientific.Reports.3}}
\label{figure:7}
\end{figure}

\subsection{Manipulation of long-range magnetic order through temperature}

The energy difference between the 0K ground state of completely compensated ferrimagnetic solution, and the polarized ferromagnetic excited state is small (Fig.~\ref{figure:1}), specially for small sized flakes. Thus, we study the effect of temperature on the long-range magnetic coupling in these nanoflakes. We consider the charge neutral bare $N$=3 nanoflake, for which the energy difference $\Delta E$ is particularly small (26 meV, Fig.~\ref{figure:3}). We perform AIMD simulation within canonical ensemble using Nos\'e-Hoover thermostat~\cite{PhysRevB.33.339, PhysRevA.31.1695} at three different temperatures, 100, 200 and 300K. We equilibrate the system for 2 ps, and then the trend in free energy difference between the FM and AFM state is observed for another 10 ps. While the nanoflake remains two-dimensional  [Fig.~\ref{figure:7} (a)] at 100K,  the structure becomes heavily distorted at higher temperatures [Fig.~\ref{figure:7} (b) and (c)], and the two-dimensional nature of the flake disappears. We, therefore, consider the nanoflake at 100K for our subsequent discussion.  We calculate the time averaged (each over 2ps) free energy for both AFM and FM structure as a function of time, and the trend in the free energy difference is shown in Fig.~\ref{figure:7}.  We observe that till $\sim$ 4 ps, the completely compensated ferrimagnetic structure is the ground state, while the ferromagnetic solution with 6$\mu_B$ total moment becomes the ground state beyond 4 ps, and remains the same for entire simulation of 12 ps. Thus, at finite temperature the ground state magnetic solution becomes ferromagnetic, which was otherwise compensated ferrimagnet at 0K. Increasing the flake size may push the transition temperature on the higher side, as the zero temperature $\Delta E$ increases with increasing flake size (Fig.~\ref{figure:1}). This result is in very good qualitative agreement with the recent experimental observation.~\cite{Scientific.Reports.3} The experiment was done for zigzag flakes with irregular shapes, and the AFM phase was found to be the ground state below 50K, which became ferromagnetic at temperatures above 80K. Although, the shape and size of the flake in the present calculation is different, the qualitative agreement with the experimental result is remarkable.

To investigate the structural evolution during the AIMD simulation at 100K, we calculate the time averaged (each over 2 ps) bond lengths. During the simulation time 4--12 ps, we find the bonds at the edge (core) to be $\sim$ 0.02 \AA~ longer (shorter) compared to the initial structure at 0K. However, we did not find any significant difference in bond distribution between FM and AFM solutions during the AIMD simulation. This indicates that the stabilization of FM state at finite temperature is due to temperature assisted overcoming of the energy barrier between FM and AFM states, and the geometric structure does not play any significant role. The magnetization density at initial and final states of the AIMD simulation (Supplementary Information)~\cite{supple} demonstrates this temperature assisted crossover of magnetic ordering.

The influence of temperature on the electron and hole doped bare nanoflake is studied via AIMD simulations at 100K. As discussed earlier, at 0K the magnetic structure is already ferromagnetic with $\Delta E$=$-$54 meV, and $-$14 meV, for hole and electron doping, respectively (Fig.~\ref{figure:3}). We find that at 100K, the magnetic structure remains ferromagnetic for both hole and electron doping, with $\Delta E$=$-$50 meV and $\Delta E$=$-$39 meV, respectively. Thus, at finite temperature the magnetic structure is ferromagnetic for carrier doped nanoflake.

\section{Summary and Discussion}

We perform density functional theory calculations supplemented with on-site Coulomb interaction to study the effect of external perturbations on the edge magnetism in hexagonal zigzag nanographene. Specially, we study the plausible manipulation of intrinsic edge magnetism via carrier doping, chemical modification at the edge, and temperature.  We find the nature of magnetic order critically depends on the carrier doping, the on-site Coulomb interaction, and chemical modification of the edges. We find the magnetic phase diagram to be complex with the appearance of various magnetic phases including long-range FM order with net magnetic moment. Interestingly, at finite temperature (100K) the FM ordering between the edges is found to be favorable, which is otherwise AFM at 0K. This observation is in qualitative agreement with the recent experimental investigation, where FM ordering is observed beyond 80K in zigzag nanographene.~\cite{Scientific.Reports.3} Although, the present study primarily concerns small sized flakes, similar qualitative results are expected for larger flakes.

We believe, the present results will help in understanding the varied experimental results on the magnetic structure of nanographene.~\cite{PhysRevLett.98.187204,PhysRevB.81.245428,Scientific.Reports.3} For example, the substrate may generate inhomogeneous strain, which in turn may modulate the local electron correlation. Similarly, the probing techniques such as scanning tunneling microscopy and scanning tunneling spectroscopy may cause a charge transfer between the tip atoms and graphene. In addition, the open edges may be susceptible to uncontrolled chemical modification. As established in the present study, all these factors will have an important consequences on the intrinsic magnetism in nanographene.    

On the other hand, in a controlled experimental setup, these findings will have important technological consequences. For device application the overall ferromagnetic ordering (both intra-edge and inter-edge) with finite total moment (5--7$\mu_B$ for the bare flakes, and $\sim$1$\mu_B$ for hydrogenated flakes) will be useful.  We find that the magnetic transitions between various phases are also associated with a nonmonotonous reduction in the gap between the highest  occupied and the lowest unoccupied molecular levels for one spin channel. This opens up a possibility for `half-metallic' FM ordering, which would trigger spin-polarized conductance, which may be useful for spintronics applications.  The phase diagrams (Fig.~\ref{figure:4} and Fig.~\ref{figure:6}) indicate that the stabilization of FM phase crucially depends on the on-site Coulomb repulsion, which is estimated to be large for graphene and derived structures.~\cite{RevModPhys.81.109, PhysRevLett.106.236805} In addition to the local Coulomb interaction, a strong nonlocal Coulomb interaction has been also proposed for graphene.~\cite{RevModPhys.81.109, PhysRevLett.106.236805} Recently a generalized Hubbard model with both local and nonlocal Coulomb interaction was mapped into an effective Hubbard model with modified local Coulomb interaction only, which is estimated to be $\sim$ 4 eV for infinite graphene layer.~\cite{PhysRevLett.106.236805, PhysRevLett.111.036601} For nanoflakes the reduction in size is expected to make electrons more localized, and thus enhancing the Coulomb interaction. Considering a local Coulomb interaction in the 4--6 eV range, a FM solution can be achieved through realistic carrier doping for both bare and hydrogenated zigzag nanographene.  The predicted quantum phase transitions between various magnetic phases can be tested experimentally by controlling carrier density through gate voltage, and tuning Coulomb interaction by inducing strain in the graphene structures by growing on different substrates.  We hope the present study will motivate further experimental investigations. 

\section{Acknowledgement}
MK acknowledges grant from the Department of Science and Technology, India under Ramanujan Fellowship. TSD thank Department of Science and Technology, India for financial support through the Unit of Nano-science and Nanotechnology, and Thematic Unit of Excellence. This research was supported in part by the Department of Science and Technology through computing resources provided by the High Performance Computing Facility at the Inter University Accelerator Centre, Delhi. We also acknowledge the supercomputing facility at the Centre for Development of Advanced Computing, Pune.

%

\end{document}